\documentclass[]{tPHM2e}

\usepackage{graphics}
\usepackage{epsfig}
\usepackage{subfigure}

\newcommand{\be}{\begin{equation}}
\newcommand{\ee}{\end{equation}}

\newcommand{\bea}{\begin{eqnarray}}
\newcommand{\eea}{\end{eqnarray}}

\newcommand{\reffig}[1]{FIG.~\ref{#1}}
\newcommand{\reftab}[1]{TABLE.~\ref{#1}}

\begin{document}



\title{Rescaled potential for transition metal solutes in $\alpha$-iron}

\author{D.J. Hepburn$^{\rm a}$$^{\ast}$\thanks{$^\ast$Corresponding author. Email: dhepburn@ph.ed.ac.uk
\vspace{6pt}}, G.J. Ackland$^{\rm a}$\vspace{6pt} and P. Olsson$^{\rm b}$\\\vspace{6pt} $^{\rm a}${School of Physics, University of Edinburgh, James Clerk Maxwell Building, King's Buildings, Mayfield Road, Edinburgh EH9 3JZ, United Kingdom};\\\vspace{6pt} $^{\rm b}${D\'{e}partement MMC, EDF-R\&D, Les Renardi\`{e}res, F-77250 Moret sur Loing, France}\\\vspace{6pt}\received{Soon}}

\maketitle

\begin{abstract}
We present empirical potentials for dilute transition metal solutes in
$\alpha$-iron. It is in the Finnis-Sinclair form and is therefore
suitable for billion atom molecular dynamics simulations. First
principles calculation shows that there are clear trends across the
transition metal series which enable us to relate the rescaling
parameters to principal quantum number and number of $d$ electrons.

The potential has been developed using a rescaling
technique to provide solute-iron and solute-solute interactions from
an existing iron potential.\bigskip

\begin{keywords}
empirical potential ; transition metal ; multicomponent ; rescaling
\end{keywords}\bigskip

\end{abstract}

\section{Introduction}

The ability to model many component alloys of iron on the atomic level
would provide an extremely powerful tool for research into the
behaviour of these materials. In particular it would allow the effects
of varying proportions of solutes on the properties of these materials
to be studied in detail. The theoretical insights gained would
complement the already extensive understanding of such systems found
from experiment and ultimately impact on the design of new materials
to suit particular applications such as those for the nuclear
industry.

In principle one would wish to model these materials using ab-initio
electronic structure calculations but such techniques are prohibitive,
being restricted to less that 1000 atoms and femto second
timescales. Alternative higher level modelling techniques exist,
such as kinetic monte-carlo (kMC) and molecular dynamics simulations
(MD) using empirical potentials, that remove these restrictions but at
the expense of requiring input from experiment or ab-initio
calculations to fix their free parameters. In particular, empirical
potentials allow billion atom simulations to be performed over nano
second timescales. The results from such simulations are readily used
as input to continuum engineering models and ultimately in the design
of new materials.

The Finnis-Sinclair scheme~\cite{FS} was based around the idea of a second
moment model to the local density of states.  In this model the band
energy depends on the width of the band, the shape of the band, and
the occupation of the band.  The moments theorem~\cite{duc} shows how
band width can be determined from the sum of squares of hopping
integrals, which forms the physical basis for the cohesive term in
Finnis-Sinclair potentials.  The band shape and occupation are
implicitly assumed to be constant.

For elements in a single phase with charge neutrality~\cite{AVF}, the
$d$-band occupation is essentially constant, and the band shape does not change
massively.  This underlies the success of single-element potentials.
Fitting to alloys has a more troubled history.  Whereas isoelectronic
alloys for isostructural elements work well~\cite{ATVF}, potentials
for systems involving structural phase transitions or elements from
different series tend to have poor transferrability from the
composition at which they are fitted.  Thus the generalization to
multicomponent alloys seems to require physics beyond the
second-moment model. Difficulties also arise in the Finnis-Sinclair
and related schemes such as the embedded atom method
(EAM)~\cite{EAM,JohnsonEAM,twoband} due to the increasing complexity
of the model. For an $N$-component system the number of parameterised
functions grows as $N^2$, as does the number of data points required
to fit the parameters of these functions.

There is one important exception where the bandshape and electron
density are reasonably constant, and we might hope that the
second-moment approach will work.  That is multicomponent alloys with
one dominant element and multiple minority elements, a particularly
relevant case here being steel. The dominant element fixes both the
crystal structure and the electron density and should, in principle,
connect the behaviour of single solute atoms to that of the dominant
element and introduce stronger connections between solute-solute
interactions and the properties of single solutes.

In this paper we present empirical potentials for single transition
metal solutes in $\alpha$-iron by rescaling the functions of a pure
iron potential. We also investigate ways to connect the interactions
between solute particles to those of single solute atoms in iron which
would allow multi-component potentials to be built once the
interactions of single solutes in iron are known. We start by
providing motivation for this procedure from the results of a recent
ab-initio study~\cite{OVD} and then give a detailed description of the
rescaling strategy. We then discuss the results for single solute
atoms in iron and present the findings of our investigation into
solute-solute interactions. Finally we present our conclusions.

\section{Rescaling}

\subsection{Ab-initio calculations: Is rescaling credible?}

If the key physics of substitutional atoms in steel is such that a
rescaling approach will work, we should expect that the rescaling will
involve the $d$-electron density and the principle quantum number.
Such an approach should work both for the perfect lattice and for
defects.  The properties of substitutional transition elements in Fe,
in particular their magnetic character, have long been known to have
systematic trends~\cite{Anisimov,Drittler}.  Here we supplement this
work with an emphasis on total energy calculations for substitutional
atoms and their interactions with point defects in bcc Fe.

We use the VASP code~\cite{vasp} with projector augmented wave (PAW)
pseudopotentials~\cite{paw} and the generalized gradient
approximation~\cite{pw} with the Vosko-Wilk-Nusair
interpolation~\cite{vwn}, which we find to give the best compromise
between computation speed and accuracy.  This gives a lattice
parameter for pure iron of 2.83\AA, which was used in the impurity and
defect calculations to define a fixed-volume supercell.  Supercells of
128±1 atoms were used with a Monkhorst-Pack 3x3x3 k-point grid
sampling the Brillouin zone. The energy cutoff was set to 300 eV. Full
details of the calculations will be published elsewhere~\cite{OVD}.

The following definition has been used for the binding energy of $n$ defects
and impurities, $\{A_i\}$: \be E_b(\{A_i\}) = \left[ \sum_{i=1}^n
E(A_i) \right] - \left[ E(\{A_i\}) + (n-1)E_0 \right],
\label{binding}
\ee where $E(A_i)$ is the energy for a configuration containing $A_i$
only, $E(\{A_i\})$ refers to a configuration containing all the
interacting entities and $E_0$ refers to a configuration containing no
defects or impurities i.e. bulk $\alpha$-iron.

\begin{figure}
\begin{center}
\subfigure[]{\label{solutevacancy}\includegraphics*[width=0.35\columnwidth]{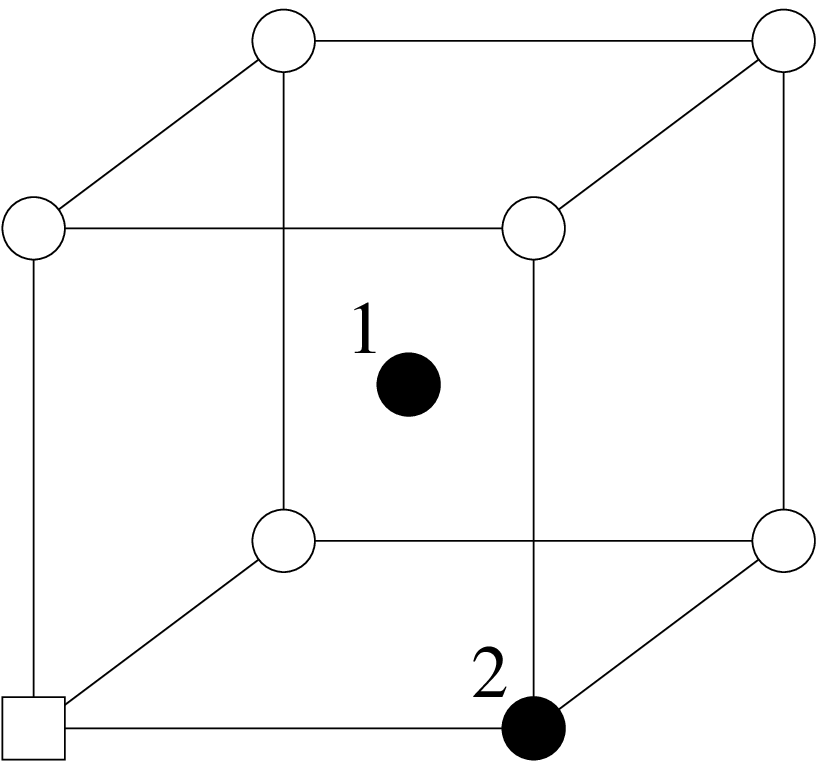}}
\subfigure[]{\label{solutesi}\includegraphics*[width=0.35\columnwidth]{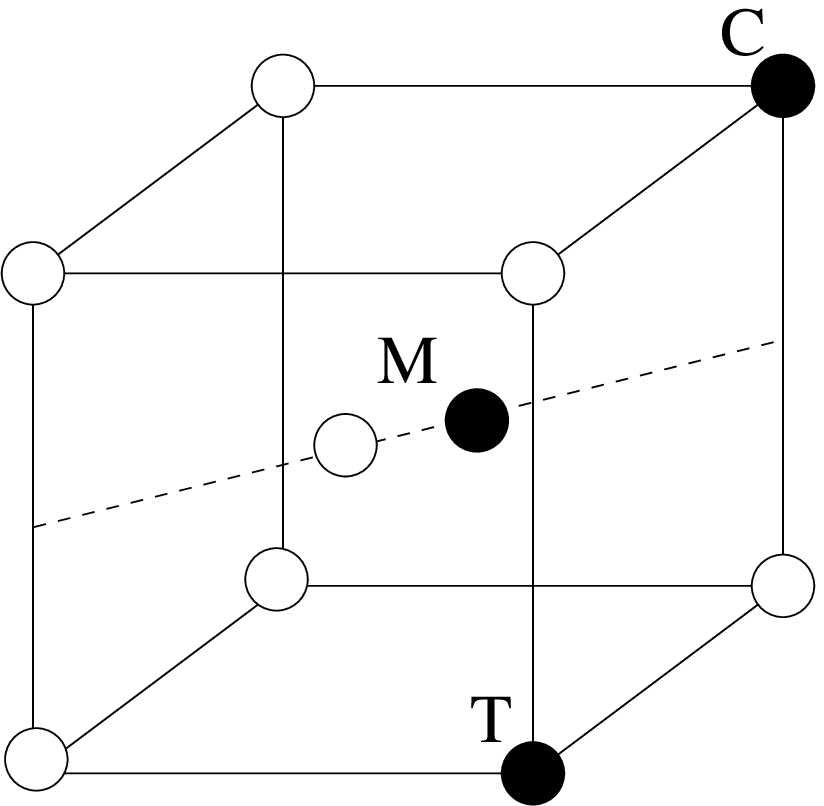}}
\caption{\label{solutedefectfig}(a) First and second nearest neighbour solute sites (black) relative to a vacancy defect (square) (b) Mixed (M), compressive (C) and tensile (T) solute sites (black) relative to a $\langle 110\rangle$-self-interstitial defect.}
\end{center}
\end{figure}

\begin{figure}
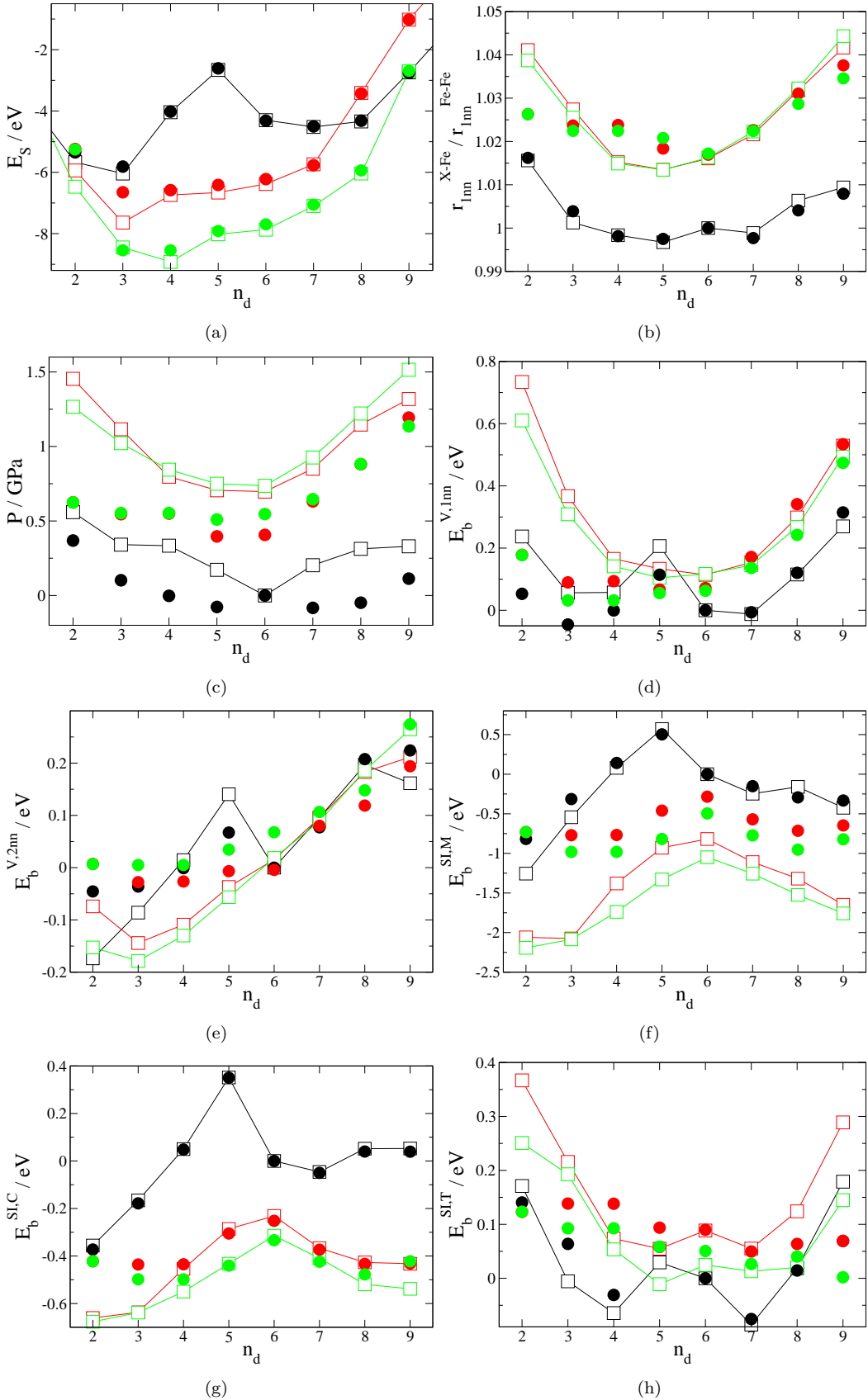

\subfigure[]{\label{esubfig}\includegraphics*[width=0.5\columnwidth]{figure2a_substitution_energies.eps}}
\subfigure[]{\label{nnsepfig}\includegraphics*[width=0.5\columnwidth]{figure2b_solute_iron_1nn_separation.eps}}
\subfigure[]{\label{pressurefig}\includegraphics*[width=0.5\columnwidth]{figure2c_solute_excess_pressure.eps}}
\subfigure[]{\label{svbindone}\includegraphics*[width=0.5\columnwidth]{figure2d_solute_vacancy_binding_1nn.eps}}
\subfigure[]{\label{svbindtwo}\includegraphics*[width=0.5\columnwidth]{figure2e_solute_vacancy_binding_2nn.eps}}
\subfigure[]{\label{ssibindm}\includegraphics*[width=0.5\columnwidth]{figure2f_solute_110sia_binding_mixed.eps}}
\subfigure[]{\label{ssibindc}\includegraphics*[width=0.5\columnwidth]{figure2g_solute_110sia_binding_compressile.eps}}
\subfigure[]{\label{ssibindt}\includegraphics*[width=0.5\columnwidth]{figure2h_solute_110sia_binding_tensile.eps}}
\caption{\label{fittargetsfig} Fit targets (squares and lines) and
corresponding values from our empirical model (circles) versus number
of d-electrons for 3d solutes (black), 4d solutes (red) and 5d solutes
(green). The first nearest neighbour solute-iron separation,
$r_\mathrm{1nn}^{X-\mathrm{Fe}}$, is plotted relative to the
corresponding iron-iron separation, $r_\mathrm{1nn}^\mathrm{Fe-Fe}$.}
\end{figure}

These total energy calculations show that there are systematic trends
across the transition metal series for the free atom substitutional
energy, $E_s$, excess pressure from a single solute, $P$, first
nearest neighbour solute-iron separation, $r_\mathrm{1nn}$,
solute-solute interactions, binding energies of a single solute to a
vacancy defect at 1nn, $E_\mathrm{b}^\mathrm{V,1nn}$, and 2nn,
$E_\mathrm{b}^\mathrm{V,2nn}$, separations and the binding energies to
a $\langle 110\rangle$-self-interstitial defect in the mixed,
$E_\mathrm{b}^\mathrm{SI,M}$, compressive,
$E_\mathrm{b}^\mathrm{SI,C}$, and tensile sites
$_\mathrm{b}^\mathrm{SI,T}$, (see~\reffig{solutedefectfig} for
configurations) as shown in~\reffig{fittargetsfig}.  The free atom
substitutional energy was calculated from our ab-initio results for
the substitution energy from the pure equilibrium phase and the
experimental cohesive energies of the pure phases~\cite{Cohesive}. We
take these values as fit targets in order to determine the parameters
of our potentials, as discussed in the following section.

We plot all energies against the number of $d$-electrons in the free
atom. In the solid this number will be affected by $s-d$ transfer of
approximately 0.5 electrons per atom.  Thus although there are clearly
different trends for more-than or less-than half filled bands,
rigorously defining which material corresponds to a half-filled
$d$-band is not straightforward.

Two elements produce outlier behaviour: chromium and manganese.  Curiously, iron
chromium and manganese were elements which Finnis Sinclair were unable to
fit with their original scheme~\cite{FS}.  These elements exhibit
unusual magnetic behaviour, which presumably accounts for this.

\subsection{Rescaling Strategy}

The starting point for our fitting strategy is the pure iron EAM
potential of Ackland {\it et al.}~\cite{ABC}. We have chosen this iron
potential over those from more recent
works~\cite{MendelevFe,AcklandFe,MullerFe} because it reproduces many
of the properties of iron despite its relatively simple form.

The most general form for the energy, $U$, of an EAM potential is
given by
\bea 
U(\{r_{ab}\}) &=& \sum_{a, b > a} V^{(X_a,X_b)}(r_{ab}) + \sum_a
F^{(X_a)} ( \rho_a ),
\label{eamequation} \\
\rho_a &=& \sum_{b\ne a}\phi^{(X_a,X_b)}(r_{ab}), \eea where
$V^{(X_a,X_b)}$, $\phi^{(X_a,X_b)}$ and $F^{(X_a)}$ are parameterised
functions dependent on the atomic species, $X_a$ and $X_b$. The
cross-species pair functions are taken to be symmetrical here,
i.e. $V^{(X,Y)} \equiv V^{(Y,X)}$ when $X \ne Y$, as are the
functions, $\phi^{(X,Y)}$.

We use the same forms for the component functions of our potential as
used in the pure iron potential~\cite{ABC}. In particular we define
the pair functions by \be V^{(X,Y)}(r) = \left\{
\begin{array}{ll}
\frac{Z_X Z_Y e^2}{4\pi \epsilon_0 r} \xi(r/r_s) & r \le r_1 \\ 
& \\ 
\exp(B_0 + B_1 r + B_2 r^2 + B_3 r^3) & r_1 < r \le r_2 \\
& \\ 
C^{(X,Y)}(r) = \sum_{k=1}^6 a_k^{(X,Y)} (r_k^{(X,Y)} - r)^3 H(r_k^{(X,Y)}-r) & r > r_2
\end{array}
\right.  
\ee
where $Z_X$ is the atomic number of species $X$, $r_s = 0.88534\ a_b /
\sqrt{ Z_X^{2/3} + Z_Y^{2/3} }$, $a_b$ is the Bohr radius and 
\be 
\xi(x) = 0.1818e^{-3.2x} + 0.5099e^{-0.9423x} + 0.2802e^{-0.4029x} + 0.02817e^{-0.2016x}. \ee

The functional form used below $r_1 = 0.9$\AA~is the universal
screened potential of Biersack and Ziegler~\cite{bz}, above
$r_2=1.9$\AA~is a parameterised cubic spline with cutoffs implemented
by the use of Heaviside step functions, $H$, and between these is an
interpolating function that ensures continuity of the function and its
derivative.

For the embedding functions, $F^{(X)}$, we take the standard square
root form for all atomic species i.e. $F^{(X)}(\rho) = -\sqrt \rho$.

The $\phi$ functions take the form of a simple cubic spline
\be
   \phi^{(X,Y)}(r) = \sum_{k=1}^2 A_k^{(X,Y)} (R_k^{(X,Y)} - r)^3
H(R_k^{(X,Y)}-r). 
\ee

For the pure iron component functions i.e. $V^\mathrm{(Fe,Fe)}$ and
$\phi^\mathrm{(Fe,Fe)}$ we take the parameters directly
from~\cite{ABC}. Iron-solute interactions are defined by rescaling
these two functions using rescale parameters, $\{p_i^{(X)}\}$:
\bea
   C^{(\mathrm{Fe},X)}(p_1^{(X)} r) & = & p_2^{(X)} C^\mathrm{(Fe,Fe)}(r) \\
   \phi^{(\mathrm{Fe},X)}(p_3^{(X)} r) & = & p_4^{(X)} \phi^\mathrm{(Fe,Fe)}(r)
\eea
This is equivalent to a direct rescaling of the parameters of the
cubic spline functions given by, for example,
\bea
    r_k^{(\mathrm{Fe},X)} & = & p_1^{(X)} r_k^\mathrm{(Fe,Fe)} \\
    a_k^{(\mathrm{Fe},X)} & = & \frac{p_2^{(X)}}{{p_1^{(X)}}^3} a_k^\mathrm{(Fe,Fe)}.
\eea
We take the rescaling factors, $\{p_i^{(X)}\}$, to be the adjustable
parameters for the purposes of fitting. The trends in the fit target
data should therefore translate to trends in these rescale parameters
across the transition metal series. In fact it should be possible to
quantify these trends by finding functional forms that relate the
rescale parameters to the elementary electronic properties of the
solutes. We present our results for the rescale parameters and a
functional form for them in terms of the number of d-electrons per
atom, $n_d^{(X)}$, in the following section.

Solute-solute interactions are defined by a similar rescaling procedure:
\bea
   C^{(X,Y)}(p_1^{(X,Y)} r) & = & p_2^{(X,Y)} C^\mathrm{(Fe,Fe)}(r) \\
   \phi^{(X,Y)}(p_3^{(X,Y)} r) & = & p_4^{(X,Y)} \phi^\mathrm{(Fe,Fe)}(r).
\eea
However, we do not determine these rescale parameters from
fitting. Instead we relate them to the rescale parameters for the
iron-solute interactions, i.e.
\be 
p_i^{(X,Y)} = p_i^{(X,Y)}(p_i^{(X)},p_i^{(Y)}).
\ee 
This is the key step that ensures we can construct multi-component
alloys once the iron-solute interactions are known.

\section{Single solute interactions in iron}

In order to determine the rescale parameters, $\{p_i^{(X)}\}$, we fit
to the ab-initio data shown in~\reffig{fittargetsfig} for transition
metal solutes in $\alpha$-iron. We do not, however, fit to
$E_\mathrm{b}^\mathrm{SI,M}$ as no satisfactory results were found
with this quantity included.

The fitting procedure was accomplished by minimising a standard least
squares response function, $\chi^2$, of the fit parameters, $\{p_i\}$,
given in terms of the fit targets, $\{t_r\}$, model values,
$\{m_r(\{p_i\})\}$, and weight factors, $\{\sigma_r\}$, by \be
\chi^2(\{p_i\}) = \sum_r \left( \frac{m_r(\{p_i\}) - t_r}{\sigma_r}
\right)^2.  \ee Our potential model values were all calculated in
atomically relaxed 4x4x4 bcc unit cell configurations, i.e. 128 atoms
before the introduction of defects and solutes, at the equilibrium
volume for the pure iron potential,
i.e. $a_0=2.8665\mathrm{\AA}$~\cite{ABC}. This was done in order to
appropriately match the ab-initio fit target data. We chose weight
factors of 0.01eV~(or $1\%$ of the fit target value if that is larger)
for energies, $0.005\mathrm{\AA}$ for lengths and $5\times
10^{-4}$eV/$\mathrm{\AA}^3$ for pressures in our fits.

\begin{figure}
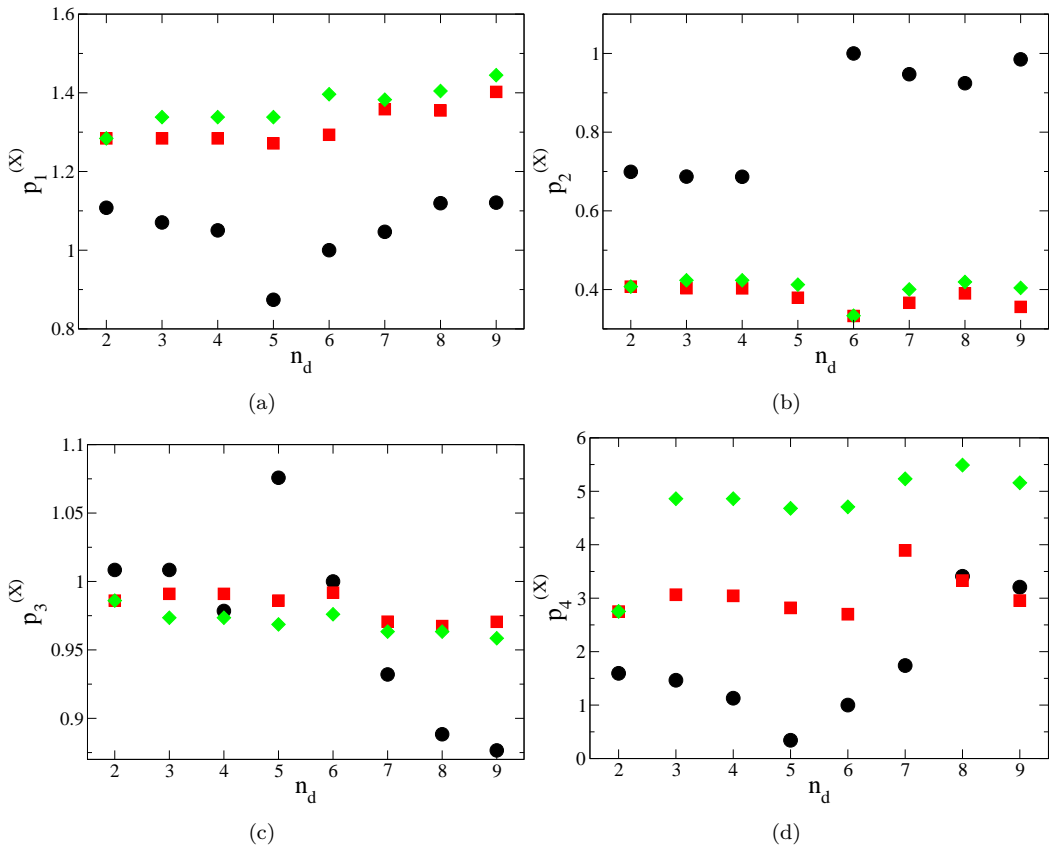

\subfigure[]{\label{pairlength}\includegraphics*[width=0.5\columnwidth]{figure3a_param_pair_length_rescale.eps}}
\subfigure[]{\label{pairvalue}\includegraphics*[width=0.5\columnwidth]{figure3b_param_pair_value_rescale.eps}}
\subfigure[]{\label{philength}\includegraphics*[width=0.5\columnwidth]{figure3c_param_phi_length_rescale.eps}}
\subfigure[]{\label{phivalue}\includegraphics*[width=0.5\columnwidth]{figure3d_param_phi_value_rescale.eps}}
\caption{\label{rescaleparamsfig} Rescale parameters as a function of the number of d-electrons for the 3d solutes (black circles), 4d solutes (red squares) and 5d solutes (green diamonds). The graphs are for (a) $p_1^{(X)}$, (b) $p_2^{(X)}$, (c) $p_3^{(X)}$ and (d) $p_4^{(X)}$.}
\end{figure}

The fitted rescale parameters are given in \reffig{rescaleparamsfig}
and \reftab{rescaleparamstab}. It it immediately clear that there are
trends across the series, especially for the 4d and 5d transition
metal solutes. The one notable exception is manganese whose anomalous
rescale parameters match the equally anomalous properties of the
solute itself. 

Looking at~\reffig{fittargetsfig} we can see that the trends in the
rescale parameters translate to the potential model values
themselves. There is especially good reproduction of the substitution
energy across all three series. The binding energy,
$E_\mathrm{b}^\mathrm{SI,C}$, is reproduced almost as well but shows
slight deviation from the ab-initio target data at the ends of the
series and especially for low $n_d$. Such deviations from the usually
parabolic trends in the ab-initio data are seen generally for the
other fit targets. The most notable is for the binding energy,
$E_\mathrm{b}^\mathrm{SI,T}$, where the potential models show
approximately linear behaviour in $n_d$. The overall deviations and
the quality of the fits is best quantified via the response function
values, as shown in~\reftab{chisqtab}. It is clear from this data that
the potentials for the low $n_d$ solutes perform especially
poorly. This is an interesting results because the influence of
s-electrons become increasingly important for these elements and their
effects are not included in the pure iron potential we have rescaled
here.

It is also clear from the response function data that the 3d solute
potentials perform better overall than those of the other two series
despite the presence of complex magnetic interactions. Even the
anomalous properties of manganese are reproduced well. The excess
solute pressures are, however, significantly underestimated although
this is true of the 4d and 5d elements also. 

Finally it is worth returning to the difficulties experienced in
fitting the binding energy of the mixed interstitial,
$E_\mathrm{b}^\mathrm{SI,M}$. As can be seen from~\reffig{ssibindm}
our potentials significantly underestimate the magnitude of this value
for the 4d and 5d elements. Including this value in the fits did
result in a more accurate reproduction but at too much cost to the
reproduction of the other fit targets. Despite this the mixed site is
still preserved as the least favoured for a solute to occupy around a
$\langle 110\rangle$-self-interstitial defect.  This failure is
unlikely to have serious consequences for any molecular dynamics
simulations using our potentials, since the mixed dumbbell site is not a migration barrier.  The interstitial is repelled by the solute by well
above any realistic thermal energy and so the mixed interstitial site will occur with very low probability.

\section{Solute-solute interactions in iron}

\begin{figure}
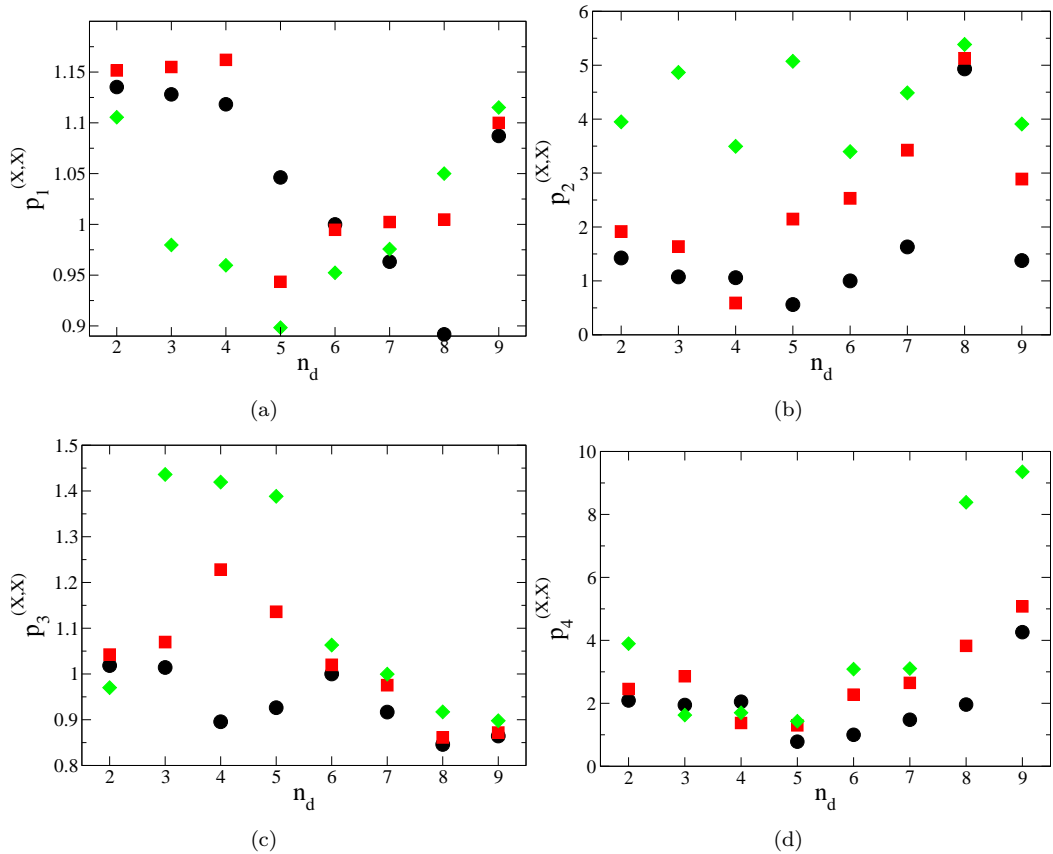

\subfigure[]{\label{pairlengthXX}\includegraphics*[width=0.5\columnwidth]{figure4a_param_pair_length_rescale_XX.eps}}
\subfigure[]{\label{pairvalueXX}\includegraphics*[width=0.5\columnwidth]{figure4b_param_pair_value_rescale_XX.eps}}
\subfigure[]{\label{philengthXX}\includegraphics*[width=0.5\columnwidth]{figure4c_param_phi_length_rescale_XX.eps}}
\subfigure[]{\label{phivalueXX}\includegraphics*[width=0.5\columnwidth]{figure4d_param_phi_value_rescale_XX.eps}}
\caption{\label{rescaleparamsXXfig} Rescale parameters as a function of the number of d-electrons for the 3d solutes (black circles), 4d solutes (red squares) and 5d solutes (green diamonds). The graphs are for (a) $p_1^{(X,X)}$, (b) $p_2^{(X,X)}$, (c) $p_3^{(X,X)}$ and (d) $p_4^{(X,X)}$.}
\end{figure}

\begin{figure}
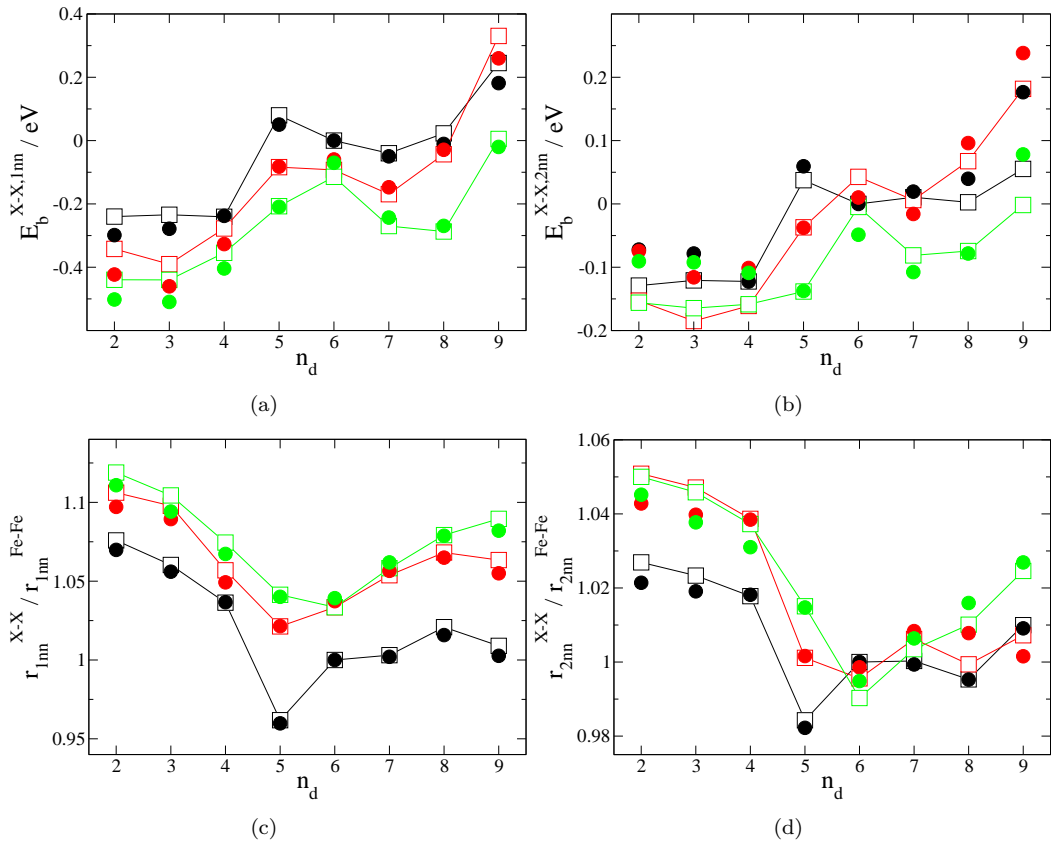

\subfigure[]{\label{essone}\includegraphics*[width=0.5\columnwidth]{figure5a_solute_solute_binding_1nn.eps}}
\subfigure[]{\label{esstwo}\includegraphics*[width=0.5\columnwidth]{figure5b_solute_solute_binding_2nn.eps}}
\subfigure[]{\label{rssone}\includegraphics*[width=0.5\columnwidth]{figure5c_solute_solute_1nn_relative_separation.eps}}
\subfigure[]{\label{rsstwo}\includegraphics*[width=0.5\columnwidth]{figure5d_solute_solute_2nn_relative_separation.eps}}
\caption{\label{fittargetsXXfig} Solute-solute fit targets (squares and lines) and
corresponding values from our empirical model (circles) versus number
of d-electrons for 3d solutes (black), 4d solutes (red) and 5d solutes
(green).}
\end{figure}

In order to gain some insight into the possible relationship between
the solute-solute rescale parameters and the solute-iron rescale
parameters we have fitted the rescale parameters, $\{p_i^{(X,X)}\}$,
to reproduce ab-initio values~\cite{OVD} for the solute-solute binding energies
from 1nn to 5nn separation, $E_\mathrm{b}^{X-X,i\mathrm{nn}}$, and to
the separations between solutes at 1nn separation,
$r_\mathrm{1nn}^{X-X}$, and 2nn separation,
$r_\mathrm{2nn}^{X-X}$. The resulting values for the rescale
parameters are shown in~\reffig{rescaleparamsXXfig} and our model
values are compared with the ab-initio fit targets
in~\reffig{fittargetsXXfig}.

The fit targets in ~\reffig{fittargetsXXfig} once again show a
two-part trend across the group, which gives us hope that not only can
rescaling be used to fit the potential functions directly, but also
that the rescale parameters can be deduced directly from the atomic
number.  The picture emerging from the fit parameters
themselves~\reffig{rescaleparamsXXfig} is less clear.  For elements
above half-filling there are clear trends with principle quantum
number and $n_d$.  However, for 5d elements with $n_d<5$ there is
considerable scatter for $p_2^{(X,X)}$ and $p_3^{(X,X)}$.  It appears
that for a longer-ranged $\phi$ can compensate for a stronger repulsion.
since this anomaly is not present in the fit targets
in~\reffig{fittargetsXXfig}, it must be an artifact of the fitting process itself.

\section{Conclusions}

In conclusion, we have advanced the hypothesis that the interactions
between transition metal atoms in iron can be described by simply
scaling the parameters of a Finnis-Sinclair model, and further that 
these scaling parameters are simply functions of the number of 
$d$-electrons and the principal quantum number.  Quantum mechanical 
calculations of interactions between solutes and point defect show 
clear trends across the group.  We have presented best-fit rescaled
potentials for all transition metal elements in an iron matrix. 

Our hypothesis is based on the notion that the environment around the
impurity is close to that in magnetic bcc iron, hence we do not expect
that the rescaled potentials will be transferrable to very different
electronic environments such as pure elements.  

\section*{Acknowledgements}
We gratefully acknowledge support from the EU FP7 GETMAT programme.

\appendices

\section{Rescale parameters for solute-iron interactions}

\begin{table}
\caption{Rescale parameters for iron-solute interactions}
\begin{tabular}{ccccc}
\toprule
Solute,$X$ & $p_1^{(X)}$ & $p_2^{(X)}$ & $p_3^{(X)}$ & $p_4^{(X)}$ \\
\hline
Ti & 1.1079352022752937 & 0.69914888963867840 & 1.0084290227415416 & 1.5946911193111377 \\
V & 1.0704687944688829 & 0.68703519774868570 & 1.0084191748980367 & 1.4652839386546592 \\
Cr & 1.0504446925177677 & 0.68660606895558630 & 0.9784540205196225 & 1.1287695242404665 \\
Mn & 0.8740657069080267 & 2.37899559818691970 & 1.0757805626557944 & 0.3418424453237209 \\
Co & 1.0467024050842750 & 0.94717978096230060 & 0.9320535734737820 & 1.7396896806120596 \\
Ni & 1.1194809316877880 & 0.92438826748289380 & 0.8883635622171985 & 3.4100636005122293 \\
Cu & 1.1208802828524023 & 0.98505124753645820 & 0.8765649836565008 & 3.2065254293566600 \\
\hline
Zr & 1.2841655075134310 & 0.40739377742609160 & 0.9860071367651349 & 2.7482154017401133 \\
Nb & 1.2841655075134310 & 0.40336016805208613 & 0.9909619464976229 & 3.0647879434909940 \\
Mo & 1.2841655075134310 & 0.40306261907883430 & 0.9909619275965241 & 3.0437240549628903 \\
Tc & 1.2714509975380506 & 0.37927270590314116 & 0.9860317313257698 & 2.8186046924493455 \\
Ru & 1.2932752539714167 & 0.33285431259332965 & 0.9918471983137220 & 2.7002031043018238 \\
Rh & 1.3583482886258411 & 0.36634462066369505 & 0.9706236772142814 & 3.8933887256877090 \\
Pd & 1.3551721040070680 & 0.39024726568702440 & 0.9674492344138608 & 3.3285717131180084 \\
Ag & 1.4023136316534160 & 0.35592505640677500 & 0.9705185530509607 & 2.9551289346288447 \\
\hline
Hf & 1.2841655075134310 & 0.40746747208599430 & 0.9860071367651349 & 2.7521572914348287 \\
Ta & 1.3381889417068988 & 0.42350948842405384 & 0.9734862986585047 & 4.8610714605878340 \\
W & 1.3381889417068988 & 0.42350948842405384 & 0.9734862986585047 & 4.8610567185373155 \\
Re & 1.3381889417068988 & 0.41247111384741486 & 0.9686430832422932 & 4.6812582679644960 \\
Os & 1.3963331257722620 & 0.33347841442944126 & 0.9760396335905974 & 4.7084791631263130 \\
Ir & 1.3825438675169937 & 0.40034598076904200 & 0.9633439996351745 & 5.2329577841196330 \\
Pt & 1.4045418753797260 & 0.41926814851469420 & 0.9633439996351745 & 5.4900896361876255 \\
Au & 1.4447702630891220 & 0.40420070742435266 & 0.9585272796369986 & 5.1583042920958950 \\
\botrule
\end{tabular}
\label{rescaleparamstab}
\end{table}

\begin{table}
\caption{Response function values, $\chi^2$, for the best rescale parameters given in~\reftab{rescaleparamstab}.}
\begin{tabular}{ccc}
\toprule
Solute,$X$ & $\chi^2(\mathrm{best})$ & \\
\hline
Ti & 568.9656946430212 & \\
V & 221.2048362240138 & \\
Cr & 99.63599215304797 & \\
Mn & 178.81158941110317 & \\
Co & 43.901666493113794 & \\
Ni & 65.72761363633806 & \\
Cu & 204.8091878556158 & \\
\hline
Zr &  4796.903769001561 & \\
Nb & 1689.7527370510506 & \\
Mo & 216.13918658727746 & \\
Tc & 137.2039869893378 & \\
Ru & 69.61819599245268 & \\ 
Rh & 26.410945708888832 & \\
Pd & 103.65177965112031 & \\
Ag & 477.68739615717334 & \\
\hline
Hf & 3491.8432724139166 & \\
Ta & 1502.974429434891 & \\
W & 408.3356703154182 & \\
Re & 198.776262870386 & \\
Os & 85.53222566998141 & \\
Ir & 35.81070622123357 & \\
Pt & 69.88189007267775 & \\
Au & 386.0176534820233 & \\
\botrule
\end{tabular}
\label{chisqtab}
\end{table}

\end{document}